\documentclass[12pt,preprint]{aastex}

\usepackage{CJK}
\usepackage{amsmath}
\usepackage{natbib}
\usepackage{longtable, lscape}
\usepackage{graphicx} 
\usepackage{epsfig}
\usepackage{xcolor}
\usepackage[colorlinks = true,linkcolor = blue,citecolor = blue]{hyperref}

\begin{document}
\begin{CJK}{UTF8}{gbsn}

\title{The Chemical Composition of an Extrasolar Kuiper-Belt-Object\footnote{ Part of the data presented herein were obtained at the W.M. Keck Observatory, which is operated as a scientific partnership among Caltech, the University of California and NASA. The Observatory was made possible by the generous financial support of the W.M. Keck Foundation.
}}

\author{S. Xu(许\CJKfamily{bsmi}偲\CJKfamily{gbsn}艺)\altaffilmark{a}, B. Zuckerman\altaffilmark{b} ,P. Dufour\altaffilmark{c}, E. D. Young\altaffilmark{d}, B. Klein\altaffilmark{b}, M. Jura\altaffilmark{b}}
\altaffiltext{a}{European Southern Observatory, Karl-Schwarzschild-Stra{\ss}e 2, D-85748 Garching, Germany, sxu@eso.org}
\altaffiltext{b}{Department of Physics and Astronomy, University of California, Los Angeles CA 90095-1562, USA}
\altaffiltext{c}{Institut de Recherche sur les Exoplan$\grave{e}$tes (iREx) and D$\acute{e}$partement de physique, Universit$\acute{e}$ de Montr$\acute{e}$al, Montr$\acute{e}$al, QC H3C 3J7, Canada; dufourpa@astro.umontreal.ca}
\altaffiltext{d}{Department of Earth, Planetary, and Space Sciences, University of California, Los Angeles, CA 90095, USA}

\begin{abstract}

The Kuiper Belt of our solar system is a source of short-period comets 
that may have delivered water and other volatiles to Earth and the other 
terrestrial planets. However, the distribution of water and other 
volatiles in extrasolar planetary systems is largely unknown. We report 
the discovery of an accretion of a Kuiper-Belt-Object analog onto the 
atmosphere of the white dwarf WD 1425+540. The heavy elements C, 
N, O, Mg, Si, S, Ca, Fe, and Ni are detected, with nitrogen observed for the first 
time in extrasolar planetary debris. The nitrogen mass fraction is 
$\sim$ 2\%, comparable to that in comet Halley and higher than in any 
other known solar system object. The lower limit to the accreted mass is $\sim$ 10$^{22}$ g, which is about one hundred thousand times the typical mass of a short-period comet. In addition, WD 1425+540 has a wide binary companion, which could facilitate perturbing a Kuiper-Belt-Object analog into the white dwarf's tidal radius.
This finding shows that analogs to objects in our Kuiper Belt exist around other stars and could be 
responsible for the delivery of volatiles to terrestrial planets beyond 
the solar system.

\end{abstract}


\section{Introduction}

Volatile elements, such as nitrogen, carbon, and oxygen are crucial for 
the chemistry of life. The prevalence of water is of particular 
importance. The distributions of water and other volatiles in our solar 
system were controlled primarily by temperatures in the protoplanetary 
disk \citep{McSweenHuss2010}. Interior of the snow line in the disk, 
water remained in the gas phase, and rocks were, consequently, relatively 
dry. Beyond the snow line, water was stable as ice, vestiges of which 
persist to this day in some asteroids, icy moons, comets, and 
Kuiper-Belt-Objects (KBOs). Even further out, nitrogen ice condensed 
into rocks and could be present in comets from the Kuiper Belt and Oort 
Cloud.

It is commonly argued that terrestrial planets, having formed inside of 
the snow line, may have accreted from dry refractory materials, and that 
water and other volatile compounds were delivered later by the impact of 
planetesimals from the asteroid belt and/or Kuiper belt 
\citep{Morbidelli2000}. The presence of KBOs around other stars has been 
inferred by excess $\sim$ 50 K blackbody radiation 
\citep[e.g.][]{Beichman2006}. However, the concentrations of water and 
other volatiles within such extrasolar minor bodies are largely unknown.
 
The sizes and masses of some individual extrasolar planets can be measured 
when combining radial velocity and transit observations 
\citep[e.g.][]{Pepe2013}.  Therefore, the bulk density of an extrasolar 
planet can be inferred, but it is difficult to find a unique solution 
for the planet's internal composition \citep{Dorn2015}. Recently, it has 
been shown that many white dwarfs are actively accreting rocky debris 
that ``pollutes" their atmospheres \citep{Zuckerman2003, Zuckerman2010, 
Koester2014a}. Typically, a white dwarf's atmosphere is dominated by 
either hydrogen or helium because heavier elements sink due to the high 
surface gravity. Heavy elements, if detected, must therefore come from 
an external source, and in many cases that source is an accretion of 
tidally disrupted rocky objects \citep{Jura2003}. As a result, 
high-resolution spectroscopic observations of polluted white dwarfs 
provide a unique way to measure the bulk compositions of extrasolar 
planetary objects \citep{JuraYoung2014}.
 
There are at least a dozen polluted white dwarfs with detailed 
measurements of accreted material \citep{JuraYoung2014}. Their abundance 
patterns are similar to those of rocky objects in our solar system, 
including bulk Earth and chondrite meteorites from the asteroid belt -- 
O, Fe, Si, and Mg are the four dominant elements in familiar 
proportions.  Volatile elements are almost always depleted, just as they 
are in Earth and meteorites from the asteroid belt \citep{Gaensicke2012, Jura2012, 
Xu2014}. Excess oxygen relative to which can be bound to 
rock-forming elements like Mg, Si, Fe, and Ca has only been detected in 
three white dwarfs. The excess oxygen is attributable to accretion of 
H$_2$O rich objects \citep{Farihi2013}. The rarity of water among the 
objects sampled by polluted white dwarfs to date is likely to be an 
indicator of low primordial water content because water and volatiles 
are generally expected to survive the post main sequence evolution of 
the host star \citep{JuraXu2010, MalamudPerets2016}.  In general, the 
overall compositions of extrasolar rocky material identified thus far 
closely resemble those of dry, relatively volatile-poor asteroids in our 
solar system.
 
Here, we report results of high-resolution spectroscopic observations of 
the polluted white dwarf WD 1425+540 with the {\it Keck Telescope} and 
the {\it Hubble Space Telescope}. WD 1425+540 (G200-39, 14:27:36.0, 
+53:48:28.4 J2000) was first reported in Greenstein's degenerate star catalog 
\citep{Greenstein1974}. It has a K dwarf companion G200-40 located 40.0 
arcsec away at 14:27:37.1, +53:47:49.2 J2000 \citep{Wegner1981}\footnote{The 
naming of these two objects is confusing in the literature and sometimes 
WD 1425+540 has been mistakenly referred to as G200-40. }. WD 
1425+540 has a helium-dominated atmosphere but also contains a large 
amount of hydrogen; it is the prototype of the ``DBA" class of white 
dwarfs that display both helium and hydrogen lines \citep{Liebert1979, 
JuraXu2012}.

\section{Observations and Data Reduction }

\subsection{High-Resolution Echelle Spectrograph (HIRES)}

WD 1425+540 was observed with the blue collimator of the HIRES \citep{Vogt1994} on the {\it Keck 
Telescope} on 2008 February 13 and 2014 May 22 (UT). The weather was 
clear on both nights. The C5 decker was used, and the resolution was 
$\sim$ 40,000. The total exposure time was 5960 s. Data were reduced 
using standard IRAF routines developed for reducing echelle data 
following \citet{Klein2010}. The final combined spectrum has an almost 
continuous coverage from 3115 to 5960 {\AA}.

\subsection{HST/Cosmic Origins Spectrograph (COS)}

WD 1425+540 was observed with the COS onboard the {\it Hubble Space Telescope} under program ID 13453 (PI: M. 
Jura) on 2014 December 8 (UT). We used the G130M grating with a central 
wavelength of 1291 {\AA}, which gives a wavelength coverage of 1130-1431 
{\AA} with a small gap in the middle. The spectral resolution was $\sim$ 
20,000. We had five consecutive orbits on this star and the total 
exposure time was 8358 s. The data were processed with the pipeline 
CALCOS 2.18.5. Following \citet{Jura2012}, around the O I triplet 1304 
{\AA} region, we used the timefilter to extract the night-time portion 
of the data to minimize the effect of geocoronal emissions.

\subsection{Abundance Analysis \label{sec:AbundanceAnalysis}}

WD 1425+540 has a trigonometric parallax and the derived distance is 58 
$\pm$ 13 pc \citep{vanAltena1995}. \citet{Bergeron2011} fitted the 
optical spectra and found the effective 
temperature T$_{eff}$ = 14,490 K, surface gravity log g = 7.95, 
and the hydrogen to helium ratio log n(H)/n(He) = -4.2. They also 
derived a spectroscopic distance of 56 pc, in good agreement with the 
parallax. However, this model cannot 
reproduce the strong asymmetric Ly $\alpha$ in the {\it COS} data \citep{Genest-Beaulieu2016}.

Ly $\alpha$ can be contaminated with interstellar absorption, which 
makes it stronger than expected from the optical hydrogen abundance. 
An interstellar hydrogen column density $\sim$ 10$^{22}$ 
cm$^{-2}$ is required to fit the red wing of Ly $\alpha$. However, the Schlegel map shows that the extinction E(B-V) is 
0.02 along the line of sight of WD 1425+540 \citep{Schlegel1998}. That 
corresponds to an upper limit of the hydrogen column 
density of 10$^{20}$ cm$^{-2}$, which is two orders of magnitude smaller 
than the amount of H required to fit Ly $\alpha$. Thus, interstellar absorption along the line of sight cannot explain the order of magnitude discrepancy between the model-derived optical 
and ultraviolet hydrogen abundances.

In order to search for a solution that fits all of the observed parameters, 
we computed a grid of models that cover T$_{eff}$ between 13,000 K and 
16,500 K, log g between 7.5 and 9.0, and log n(H)/n(He) between -3.0 and 
-5.0. We compared the models with the observed SED, He lines, Balmer 
lines, and Ly $\alpha$. However, we were unable to find a solution that 
fits all of the observables; the discrepancy between Balmer lines and Ly 
$\alpha$ is always present. WD 1425+540 represents an extreme case, but 
discrepancies between optical and ultraviolet hydrogen abundances have 
been observed in other helium-dominated white dwarfs 
\citep[e.g.][]{Jura2012}. \citet{Genest-Beaulieu2016} investigated this 
issue and found it is possibly due to atmospheric stratification. Finding the best white 
dwarf parameters that fit all available data is beyond the scope of this 
paper, and we defer it to a future study. Fortunately, the relative element abundances are not sensitive to white dwarf parameters for white dwarfs like WD 1425+540; they are comparable within the measurement uncertainties \citep{Klein2010, Klein2011}.

\begin{figure}[hp]
\epsscale{1.0}
\plotone{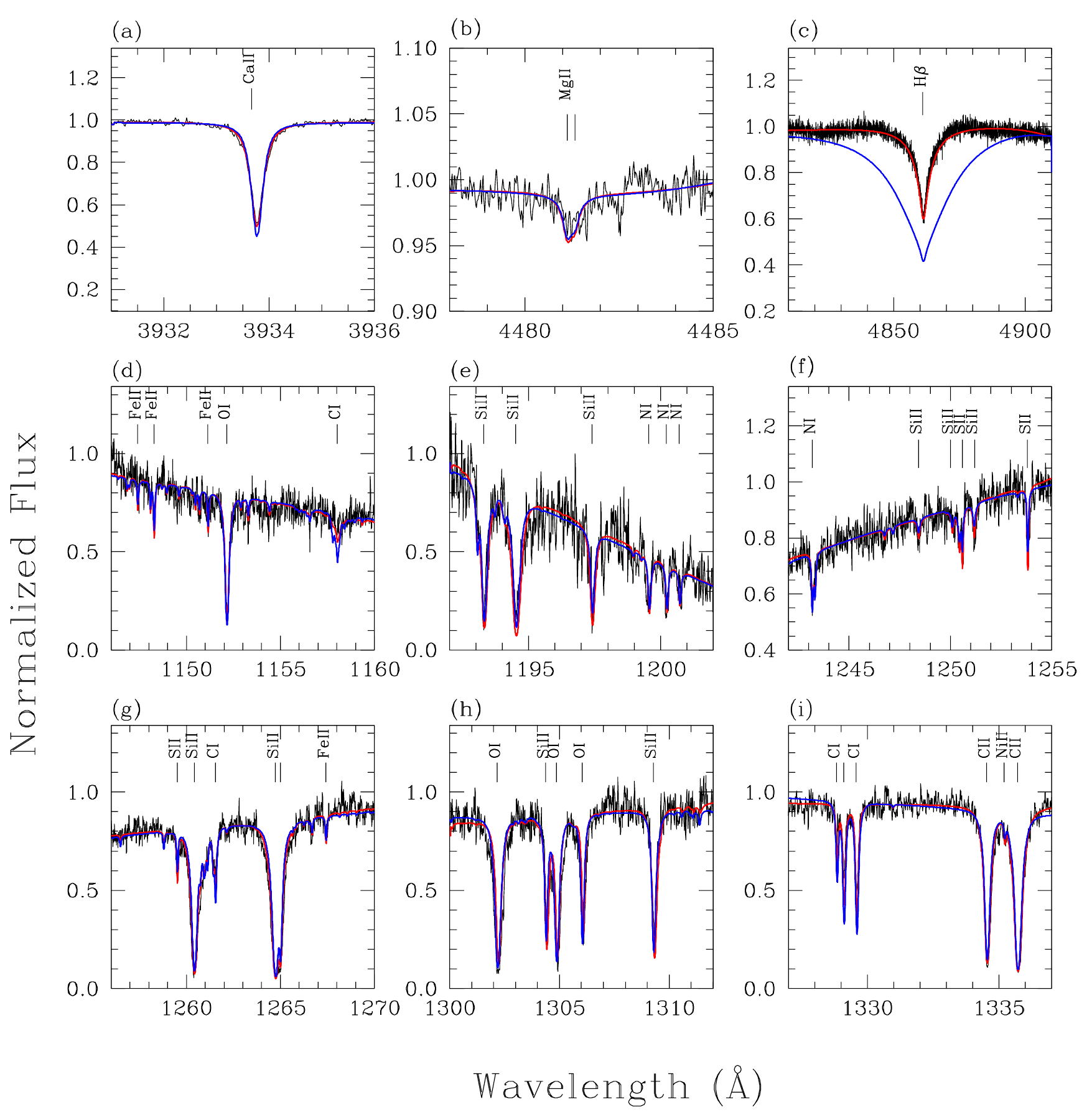}
\caption{ Representative spectra for WD 1425+540. The black line represents the data, while the red (model I) and blue (model II) lines are our model fit with abundances from Table \ref{Tab:Abundances}. The top panels are from Keck/HIRES, while the rest are from HST/COS. Some weak Fe II lines are not labeled.
}
\label{Fig:WD1425}
\end{figure}

We adopted the temperature and surface gravity as 
listed in \citet{Bergeron2011}. To assess the effect of hydrogen 
abundance on the heavy elements, we calculated synthetic white dwarf 
model spectra for two sets of hydrogen abundances: model I adopts log 
n(H)/n(He) = -4.2, as derived from Balmer lines, while model II uses log 
n(H)/n(He) = -3.0 from Ly $\alpha$\footnote{
\citet{Genest-Beaulieu2016} considered a pure helium atmosphere and 
derived log n(H)/n(He) = -2.63 for Ly $\alpha$, which is different from 
our value of -3.0. It is likely due to the presence of heavy elements, 
which changes the temperature and pressure profile of the atmosphere 
\citep[e.g.][]{Dufour2010}.}. To derive the abundances of heavy 
elements, we followed the method outlined in \citet{Dufour2010, 
Dufour2012} and divided the spectra into 10-20 {\AA} segments. Each 
element typically has 2-4 segments and each segment was fitted 
separately, as shown in Figure \ref{Fig:WD1425}. The flux level of the 
model in each segment was another free parameter; 
this was particularly important for the lines in the wings of Ly $\alpha$ 
(e.g., Figure \ref{Fig:WD1425}, panels (d), (e), (f), and (g)), because we were not able to 
reproduce the asymmetric profile of Ly $\alpha$. We used a chi-squared 
minimization routine to iterate our calculation of the model atmosphere 
until the difference between the observed spectra and the model has 
reached the minimum. The uncertainty was calculated by taking the 
standard deviation of the abundances derived from each segment. Some 
representative spectra are shown in Figure \ref{Fig:WD1425}. Both the 
{\it COS} data and the model spectra are available in the supplementary 
material.

At 58 pc, there can still be a small amount of 
interstellar absorption present in the 
ultraviolet spectra of WD 1425+540. The average radial velocity of the 
stellar spectral lines is 3.0 $\pm$ 4.8 km s$^{-1}$, very close to the 
velocity of typical interstellar lines. In order to 
assess the potential influence of interstellar lines, we only fit lines that are 
most likely to have an interstellar contribution based on previous 
studies \citep{Jura2012}; these include Si II 1260.4 {\AA}, C II 1334.5 
{\AA}, O I 1302.2 {\AA}, and the N I triplet around 1200 {\AA}. We found 
the derived abundances based on these lines to be comparable to those 
from the other photospheric lines. We conclude that interstellar lines 
do not contribute significantly to the spectra, and we used all available 
lines, as also seen in previous studies \citep{Gaensicke2012, Jura2012}, to derive the final abundances reported in Table 
\ref{Tab:Abundances}. Specifically, the abundance of N is derived from N 
I 1243.2 {\AA} (panel (f) in Figure \ref{Fig:WD1425}), the N I triplet 
around 1200 {\AA} (panel (e) in Figure \ref{Fig:WD1425}), and the N I 
triplet around 1145 {\AA}. 

\begin{table}[hp]
\begin{center}
\caption{WD 1425+540 Abundances}
\begin{tabular}{lcclcccc}\label{Tab:Abundances}
\\
\hline \hline
Element & \multicolumn{2}{c}{log n(Z)/n(He)$^a$} 	& log t$^b$(yr)&  \multicolumn{2}{c}{M$^{c}$ (10$^{20}$ g)}& \multicolumn{2}{c}{$\dot{M}$$^{d}$ ( 10$^7$ g s$^{-1}$)}\\
	& I	& II	& I\&II	& I & II & I & II\\
  \hline

H	& -4.20$\pm$0.10 	&-3.00$\pm$0.10	& ...		& 1130	& 1790	& ...	&...\\
C	& -7.29$\pm$0.17	& -7.28$\pm$0.31	& 6.164	& 11.0	& 11.4	& 2.41	& 2.48	\\
N	& -8.09$\pm$0.10	& -7.97$\pm$0.15	& 6.123	& 2.06	& 2.67	& 0.50	& 0.64 	\\
O	& -6.62$\pm$0.23	& -6.51$\pm$0.22	& 6.113	& 68.5	& 89.2	& 16.7	& 21.8 	\\
Mg	& -8.16$\pm$0.20	& -8.11$\pm$0.20	& 6.115	& 2.99	& 3.32	& 0.73	& 0.81 	\\
Si 	& -8.03$\pm$0.31	& -7.87$\pm$0.26	& 6.106	& 4.66	& 6.72	& 1.16	& 1.67 	\\
S	& -8.36$\pm$0.11	& -8.13$\pm$0.16	& 6.032	& 2.48	& 4.26	& 0.73	& 1.25 	\\
Ca	& -9.26$\pm$0.10	& -9.31$\pm$0.10	& 5.935	& 0.40	& 0.35	& 0.15	& 0.13 	\\
Fe	& -8.15$\pm$0.14	& -7.99$\pm$0.10	& 5.925	& 7.02	& 10.3	& 2.65 	& 3.90	\\
Ni	& -9.67$\pm$0.20	& -9.52$\pm$0.20	& 5.940	& 0.28	& 0.32	& 0.08 	& 0.12	\\
\\
total$^e$	&	& & &99.4	& 128.6	& 25.1 & 32.8\\
\hline
\end{tabular}
\end{center}
{\bf Note.} We present results for two models I \& II (see section \ref{sec:AbundanceAnalysis}). \\
$^a$ Measured abundances in the convective zone. The mass of the convection zone relative to the total mass of the white dwarf, q = log M$_{cvz}$/M$_{WD}$ = -5.495. \\
$^b$ Settling times out of the convective zone \citep{Dufour2016}. \\
$^c$ The total mass of each element that is currently in the convective zone.\\
$^d$ The average accretion rate of each element, calculated by dividing the mass of an element M by its settling time t.\\
$^e$ Only the heavy elements are included in this row. H is excluded.\\
\end{table}

To derive the intrinsic abundances in the accreting material, we need to know the stage of the accretion process. The accretion of material onto a white dwarf has been classified into three stages, the build-up phase, the steady-state stage, and the dissipating stage \citep{Koester2009a}. The steady-state accretion is sometimes accompanied by the presence of infrared excess from a circumstellar dust disk. The dissipating stage often shows an under-abundance of heavier elements, such as Ca, Fe and Ni, due to their relatively short settling times\footnote{For example, see Table \ref{Tab:Abundances} for settling times of different elements and \citet{JuraXu2012} for discussion about dissipating stage accretion for GD 61.}. WD 1425+540 does not have an infrared excess and the abundance ratios between Ca, Fe, and Ni relative to Mg are ``normal", compared to other rocky objects. We derive the compositions of the accreting material in both the build-up and steady-state stages, as shown in Table \ref{Tab:MassFraction}. Our main conclusion of accretion from a volatile-rich object is not dependent on the assumption of a particular accretion stage, because the difference in settling times is, at most, a factor of 1.7 (between C and Fe, see Table \ref{Tab:Abundances}).

\begin{table}[hp]
\begin{center}
\caption{Mass Fractions of Heavy Elements in Different Rocky Objects}
\begin{tabular}{ccccccllll}
\\
\hline \hline
Element & \multicolumn{4}{c}{WD 1425+540}	& Comet Halley	& Bulk Earth	& CI chondrites \\
		& I. bd	& I. std & II. bd & II. std\\
  \hline
C	& 11.2\%	& 9.6\%	& 8.9\%	& 7.6\%	& 25.8\%	& 0.17-0.39\%	& 3.2\% \\
N	& 2.1\%	& 2.0\%	& 2.1\%	& 1.9\%	& 1.6\%	& 1.2e-6	& 0.15\% \\
O	& 68.9\%	& 66.7\%	& 69.4\%	& 66.5\%	& 37.6\%	& 32.4\%	& 46.0\% \\
Mg	& 3.0\%	& 2.9\%	& 2.6\%	& 2.5\%	& 6.3\%	& 15.8\%	& 9.7\% \\
Si	& 4.7\%	& 4.6\%	& 5.2\%	& 5.1\%	& 13.7\%	& 17.1\%	& 10.5\%	\\
S	& 2.5\%	& 2.9\%	& 3.3\%	& 3.8\%	& 6.1\%	& 0.46\%	& 5.9\% \\
Ca	& 0.4\%	& 0.6\%	& 0.3\%	& 0.4\%	& 0.7\%	& 1.62\%	& 0.92\% \\
Fe	& 7.1\%	& 10.5\%	& 8.0\%	& 11.9\%	& 7.7\%	& 28.8\%	& 18.2\% \\
Ni	& 0.2\%	& 0.3\%	& 0.3\%	& 0.4\%	& 0.6\%	& 1.69\%	& 1.07\% \\
\hline
\label{Tab:MassFraction}
\end{tabular}
\end{center}
{\bf Note.} \\
The mass fraction of WD 1425+540 is shown assuming build-up (bd),  steady-state (std) accretion, log n(H)/n(He) = -4.2 (model I), and -3.0 (model II), respectively. H is excluded in the calculations because much or most of the hydrogen may not be associated with the 
current accretion event \citep{JuraXu2012}. If we take the amount of hydrogen that could have been bound with the excess oxygen (55\% by number, see section \ref{Discussion}), in the form of H$_2$O, then the 
mass fraction of hydrogen would be $\sim$ 5\%, which is comparable to 
that in comet Halley 5\% \citep{Jessberger1988}. In that case, the 
mass fraction of other heavy elements would be slightly lower. The mass 
fractions of different elements, excluding H, for a few solar system objects are also listed \citep{Jessberger1988, Allegre2001, WassonKallemeyn1988}.
\end{table}

\section{Discussion \label{Discussion}}
\begin{figure}[hp]
\plotone{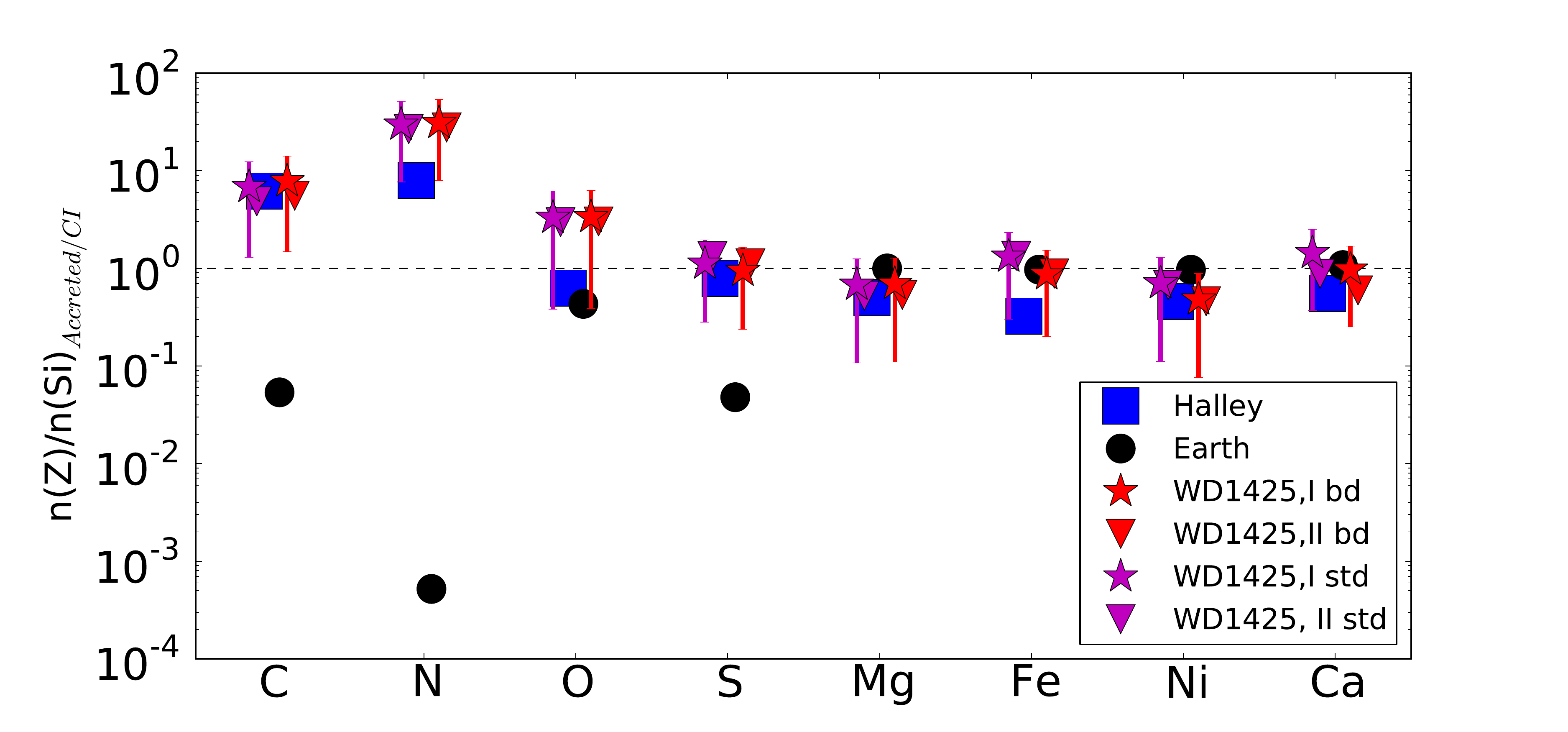}
\caption{Number fraction of various elements relative to silicon, a major constituent of rocky planetary material in our solar system. The ratios have been normalized to that of CI chondrites representing standard solar-system abundances of rock-forming elements. Four models are shown for the composition of the material accreting onto WD 1425+540; model I for log n(H)/n(He) = -4.2, model II for log n(H)/n(He) = -3.0, the build-up accretion model (bd) and the steady-state accretion model (std, see Table \ref{Tab:MassFraction}). For clarity, error bars are not shown for model II but they are comparable to those in model I. In all cases, the composition of the material accreting onto WD 1425+540 resembles that of comet Halley. \label{Fig:Halley}}
\end{figure}

Tables \ref{Tab:Abundances} and \ref{Tab:MassFraction} provide abundances and mass fractions for C, N, O, Mg, Si, S, Ca, Fe, and Ni. The overall abundance pattern derived from these data resembles the composition of comet Halley, as shown in Figure \ref{Fig:Halley}. The volatile elements nitrogen, carbon, and oxygen are enhanced relative to those in bulk Earth and CI chondrites. The relative abundances of rock-forming elements in WD1425+540, including Si, Mg, Ca, Ni, and Fe, are similar to chondritic rocks of our solar system. The total mass of heavy elements in the convective zone of WD 1425+540 is $\sim$ 10$^{22}$ g; this is about 0.10\% of the mass of the most massive KBO Pluto \citep{Buie2006}. The mass in the convective zone represents a lower limit to the total accreted mass, and the accreting object could have been more massive. 

\begin{figure}[hp]
\epsscale{1.2}
\plottwo{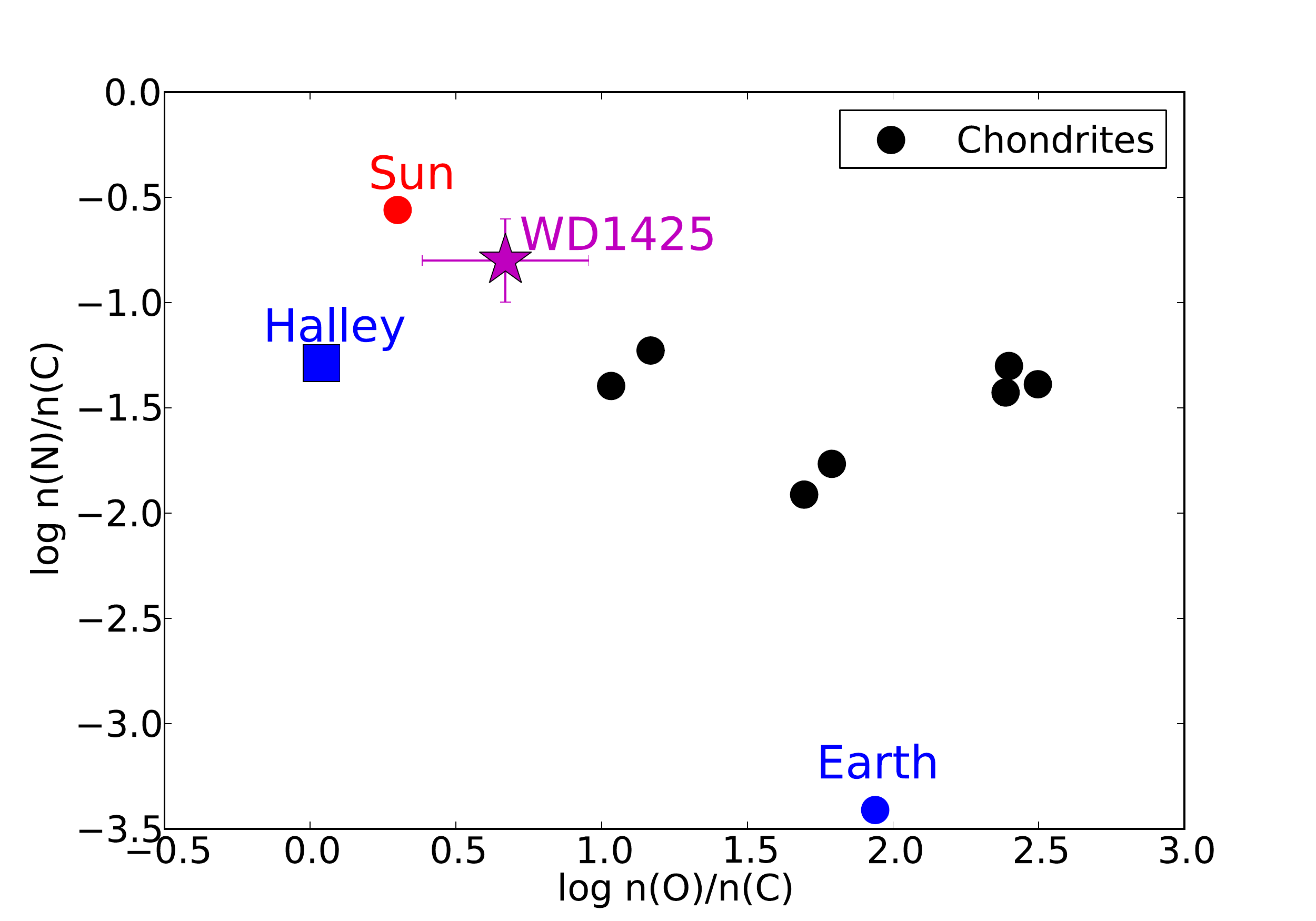}{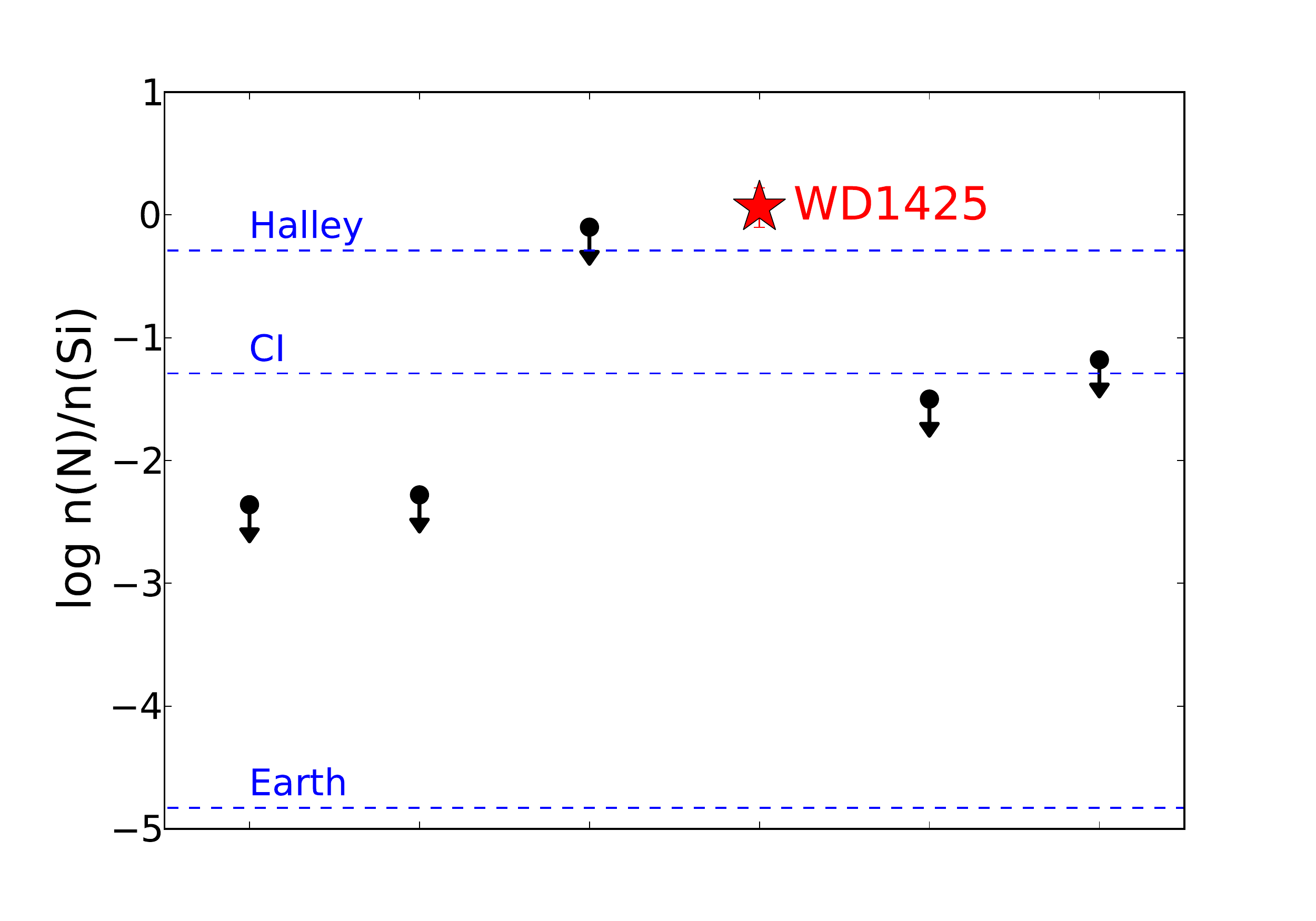}
\caption{Left panel: O/C vs N/C ratios by number for chondrites \citep{WassonKallemeyn1988}, the Sun \citep{Lodders2003}, comet Halley \citep{Jessberger1988}, bulk Earth \citep{Allegre2001}, and WD 1425+540. The abundance ratios from different models of WD 1425+540 all lie within the error bars. WD 1425+540 has accreted materials with abundance ratios of volatile elements similar to those of the Sun and comet Halley, which are different from rocky objects in the solar system \citep{JuraYoung2014}. Right panel: black dots are upper limits to N/Si ratios by number for some polluted white dwarfs. The red star represents the value for WD 1425+540, which is a measurement reported in this paper. The ratios for comet Halley, CI chondrites, and bulk Earth are also marked. WD 1425+540 has a very high nitrogen abundance that is comparable to that in comet Halley.
}
\label{Fig:Ratios}
\end{figure}

The solar-like abundance of nitrogen for the object accreted onto WD 1425+540 is higher than existing quantitative estimates of N in chondrites and Earth (see Figure \ref{Fig:Ratios}). Bulk nitrogen concentrations are largely unknown for outer solar system bodies, but nitrogen ice is prevalent on the surface of cold KBOs, including Pluto, Charon, and Eris due to the stability of N$_2$ and ammonia ices at cold temperatures \citep{Cruikshank2015}. We conclude that the abundant nitrogen present in WD 1425+540 is a signature of formation beyond the nitrogen ice line in a location analogous to our Kuiper Belt. This first detection of nitrogen in a white dwarf atmosphere provides the first estimate for the bulk nitrogen composition of a rocky/icy planetesimal anywhere, including in our own solar system. Support for the inference that we are measuring a bulk nitrogen concentration comes from the solar-like N/C ratio, suggesting little selective loss of volatiles during the formation and subsequent evolution of the accreted object \citep{Bergin2015}.  

The material accreted onto WD 1425+540 is rich in carbon as well, as shown in Table \ref{Tab:MassFraction}. Indeed, carbon is among the most abundant elements comprising the accreted minor planet. Carbon and oxygen are the most abundant heavy elements in planet-forming regions and C/O ratios can influence planetesimal geology. The C/O ratio in WD 1425+540 is 0.15-0.22 by number, indicating that the geology of this planetesimal was dominated by magnesium silicates \citep{Pontoppidan2014}. So far, no extrasolar planetary material with carbon-dominated geology has been detected \citep{Wilson2016}.
 
Oxygen is the most abundant element in the accreting material. There is an excess of oxygen relative to that which could be bound with the major rock-forming elements, including Mg, Si, Ni, Ca, and Fe. This oxygen excess suggests that the polluting object was not only rich in N and C, presumably as ices, but also in water ice. This is further supported by the large amount of hydrogen in the atmosphere. The amount of excess oxygen is $\sim$ 55\% by number. If all of the excess oxygen is attributable to H$_2$O, it would account for log n(H)/n(He) = -6.6 and the mass fraction of water ice in the object would have been about 30\%. 
 
A comet accretion model for white dwarfs was first proposed in 
\citet{Alcock1986} and more recently discussed in \citet{Bonsor2011} and 
\citet{Stone2015}. However, comet models suffer from various problems 
\citep{Zuckerman2003}, including derived abundances of heavy elements 
incompatible with observations. The mass of hydrogen in helium-dominated 
white dwarfs increases with the white dwarf cooling age, which has been 
attributed to the accretion of comet-like H$_2$O rich objects 
\citep[e.g.][]{Veras2014}. However, the increase of hydrogen abundance 
could also come from some unexplored evolutionary sequence, such as 
convective mixing of a thin hydrogen-rich atmosphere below the helium 
convection zone \citep{Dufour2007}. 

The accreting material observed in 
WD 1425+540 provides direct evidence for the presence of KBO analogs 
around stars other than the Sun. In addition, WD 1425+540 has a K dwarf 
companion at 40.0 arcsec (2240 au) away \citep{Wegner1981}. The presence 
of a distant stellar companion can impact the stability of extended 
planetary systems and, thus, enhance the chances of perturbing objects -- 
that initially orbit far from a white dwarf -- into its tidal radius via the Kozai-Lidov mechanism
\citep{Zuckerman2014, BonsorVeras2015, Naoz2016}.

When on the main sequence, WD 1425+540 would have had a mass of about 2 M$_{\odot}$ and a luminosity of 10 L$_\odot$. Thus, the 
semimajor axes of cold nitrogen-bearing KBOs
would have been about 120 au -- that is about 3 times further from the 
star than is the Sun's KBO.  Then, mass loss on the AGB would 
have moved the star's Kuiper Belt out another factor of about 3.  Therefore, 
the object that has been accreted by WD 1425+540 likely was orbiting the 
white dwarf out beyond 300 au before it began its fateful journey toward 
oblivion.

 \section{Perspective}
 
Previous studies show that white dwarfs are mostly accreting from dry, volatile-depleted minor planets. The discovery of a Kuiper-Belt-Object analog, which has polluted the atmosphere of WD 1450+540, provides strong evidence that volatile-rich planetesimals also exist around other stars. In the solar system, significant amounts of volatile compounds and water are present on KBOs, and they could have been important for delivery of these materials to Earth \citep{Morbidelli2000}. With this new dataset, we can conclude that extrasolar terrestrial planets could have volatile element and water abundances provided by KBO analogs that are comparable to those on Earth.

\end{CJK}

Acknowledgments: This paper is dedicated to the memory of Michael 
Jura, who was a pioneer in the study of polluted white dwarfs.  We thank the anonymous referee for helpful suggestions that improved the paper. We thank 
Edwin A. Bergin, Amy Bonsor, Cynthia Genest-Beaulieu and Pierre Bergeron for interesting discussions. Parts of the data reported here were obtained with the Keck 
Telescope. The authors wish to recognize and acknowledge the very 
significant cultural role and reverence that the summit of Mauna Kea has 
always had within the indigenous Hawaiian community. This research was 
supported in part by NASA grants to UCLA.

\bibliographystyle{apj}
%\bibliography{apj-jour,WD.bib}

\end{document}